\documentclass[%
 reprint,
superscriptaddress,
 amsmath,amssymb,
 longbibliography,
 aps,
]{revtex4-1}

\usepackage{soul}
\usepackage{verbatim}
\usepackage{graphicx}%
\usepackage{color}
\usepackage{dcolumn}%
\usepackage{bm}%

\usepackage{physics}
\newcommand{\xj}{{\mathcal{X}_{j}}}
\newcommand{\xk}{{\mathcal{X}_{k}}}
\newcommand{\br}{\mathbf{r}}
\newcommand{\by}{\mathbf{y}}

\newcommand{\bx}{\mathbf{x}}

\newcommand{\id}{\mathrm{d}}
\newcommand{\dbr}{\id\br}
\newcommand{\drhat}{\id\hat{R}}

\newcommand{\calx}{\mathcal{X}}

\newcommand{\KK}{\mathbf{K}}

\def\CA{{\mathcal{A}}}
\def\CB{{\mathcal{B}}}
\def\CC{{\mathcal{C}}}
\def\loss{\ell{}}
\newcommand{\Cnabla}{\nabla{\mskip-\thinmuskip}}

\newcommand{\yy}{\mathbf{y}}

\def\DD{\mathbf{D}}
\def\T{\top}

\usepackage[oldsyntax]{stackengine}
\def\LL{\stackengine{\Sstackgap}{$\mathbf{L}$}{\(\hspace{2.0pt}\hat{}\)}{O}{l}{F}{T}{S}}

\usepackage{etoolbox}
\usepackage[]{nomencl}   
    \makenomenclature

\providetoggle{nomsort}
\settoggle{nomsort}{true} %

\makeatletter
\iftoggle{nomsort}{%
    \let\old@@@nomenclature=\@@@nomenclature        
        \newcounter{@nomcount} \setcounter{@nomcount}{0}%
        \renewcommand\the@nomcount{\two@digits{\value{@nomcount}}}%
        \def\@@@nomenclature[#1]#2#3{%
          \addtocounter{@nomcount}{1}%
        \def\@tempa{#2}\def\@tempb{#3}%
          \protected@write\@nomenclaturefile{}%
          {\string\nomenclatureentry{\the@nomcount\nom@verb\@tempa @[{\nom@verb\@tempa}]%
          \begingroup\nom@verb\@tempb\protect\nomeqref{\theequation}%
          |nompageref}{\thepage}}%
          \endgroup
          \@esphack}%
      }{}
\def\three@digits#1{\ifnum#1<100 0\ifnum#1<10 0\fi\fi\number#1}
\makeatother

\begin{document}

\title{
    Machine-learning of atomic-scale properties based on physical principles
}%

\author{G\'abor Cs\'anyi}
\affiliation{
 Engineering Laboratory, University of Cambridge, Trumpington Street, Cambridge, CB2 1PZ, United Kingdom 
}%

\author{Michael J.~Willatt}
\affiliation{Laboratory of Computational Science and Modelling, Institute of Materials, \'Ecole Polytechnique F\'ed\'erale de Lausanne, Lausanne, Switzerland}%

\author{Michele Ceriotti}
\affiliation{Laboratory of Computational Science and Modelling, Institute of Materials, \'Ecole Polytechnique F\'ed\'erale de Lausanne, Lausanne, Switzerland}%

\date{\today}%

\begin{abstract}
We briefly summarize the kernel regression approach, as used recently in materials modelling, to fitting functions, particularly potential energy surfaces, and highlight how the linear algebra framework can be used to both predict and train from linear functionals of the potential energy, such as the total energy and atomic forces. 
We then give a detailed account of the Smooth Overlap of Atomic Positions (SOAP) representation and kernel, showing how it arises from an abstract representation of smooth atomic densities, and how it is related to several popular density-based representations of atomic structure.  We also discuss recent generalisations that allow fine control of correlations between different atomic species, prediction and fitting of tensorial properties, and also how to construct structural kernels---applicable to comparing entire molecules or periodic systems---that go beyond an additive combination of local environments.\footnote{Published as Ceriotti M., Willatt M.J., Csányi G. (2018) Machine Learning of Atomic-Scale Properties Based on Physical Principles. In: Andreoni W., Yip S. (eds) Handbook of Materials Modeling. Springer, Cham} 
\end{abstract}

\pacs{Valid PACS appear here}%
\maketitle

\section{Introduction}

There has been a surge of activity during the last couple of years in applying machine learning methods to materials and molecular modelling problems,
that was largely fuelled by the evident success of these techniques in what can loosely be called artificial intelligence. 
These successes have followed from the collective experience that the scientific community has gained in fitting high volumes of data with very complex functional forms that involve a large number of free parameters, while still keeping control of the regularity and thus avoiding catastrophic overfitting. In the context of molecular modelling, empirical fitting of potential energy surfaces has of course been used for many decades. Indeed it is generally held that this is the only practical way to simulate very large systems (many thousands of atoms) over long time scales (millions of time steps)\cite{Finnis:2004da}. 

Traditionally, when fitting empirical models of atomic interactions, regularity was ensured by writing functional forms that are expressed in terms of one dimensional functions, e.g. pair potentials, spherically symmetric atomic electron densities, bond orders (as a function of number of neighbours), etc. Such functions are easy to inspect visually to ensure that they are physically and chemically meaningful, e.g. that pair potentials go to zero at large distances and are strongly repulsive at close approach, that atomic electron densities are decreasing with distance, that electron density embedding functions are convex, etc. Moreover, these natural properties are easy to build into the one dimensional functional forms or enforced as constraints in the parameter optimisation. It is widely held that employing such ``physically meaningful'' functional forms are key to achieving good {\em transferability} of the empirical models\cite{Brenner:2000uh}. 

It is also recognised, however, that  the limited functional forms that can be built from these one dimensional functions ultimately limit the accuracy that these empirical models can achieve. In trying to replace them by high dimensional fits using much more flexible functional forms, two things immediately have to change. The first is the target data. When fitting only a few parameters, it is natural to demand that important observables that are deemed to be central to the scientific questions being addressed are reproduced correctly, and it is easiest to do this if they are part of the fit: e.g. melting points and other phase boundaries, radial distribution functions, etc. But in the case of very many parameters, their optimisation also takes a significant number of evaluations, and it becomes impractical to use complex observables as targets. Moreover, there is a drive towards using a ``first principles'' approach, i.e. that the potentials should actually reproduce the real Born-Oppenheimer potential energy surface with sufficient accuracy {\em and therefore} the scientifically relevant observables also. The hope is that this will result in transferability in the sense that a wide array of macroscopic observables will be correctly predicted without any of them being part of the fit explicitly, {\em and therefore}, the corresponding microscopic mechanisms that are also dependant on the potential energy surface will also be correct. 
So it is natural to take  values of the potential energy, computed by the electronic structure method of choice, as the target data. The large number of free parameters can then easily be counterbalanced by a large amount of calculated target data. 

The second thing that has to change is how the smooth physically meaningful behavior of the potential is controlled. It is not practical to inspect manually high dimensional functions to ensure that their predictions are physically and chemically meaningful for all possible configurations. Therefore it becomes even more important to build into the  functional forms as much prior information as possible about limiting behaviour and {\em regularity} (the technical word for the kind of smoothness we are interested in). Reviewing recent work, this paper sets out an example framework for how to do this. The key goals are to create functional forms that preserve the (i) invariance of the properties over permutation of like atoms, (ii) invariance of scalar and covariance of tensorial properties with three dimensional rotations, (iii) continuity and regularity with respect to changes in atomic coordinates, including compact support of atomic interactions by including finite cutoffs. 

Evidence is accumulating  that strictly enforcing these physically-motivated properties  is enormously beneficial, and many of the most successful machine-learning schemes for atomic-scale systems are built around symmetry arguments. One possible approach is to describe the system in terms of internal coordinates -- that automatically satisfy  rotational invariance -- and then symmetrize explicitly the vector of representations or the functional relation between the representations and the properties. Permutationally-invariant polynomials are an example that have been very effective to model the potential energy surfaces of small molecules (see e.g. the work of Bowman and Braams\cite{Braams:2009eb}). Sorting the elements of the representation vector according to interatomic distances has also been used as a way of obtaining permutation invariance at the cost of introducing derivative discontinuities~\cite{rupp+12prl,fabe+15ijqc,zhan+18prl}.
Another possibility, which we will focus on in this paper, starts from a representation of each structure in terms of atomic densities -- that are naturally invariant to atom permutations -- and then builds a representation that is further invariant to translations and rotations also. 

Either way, once an appropriate description of each structure has been obtained, further regularization can be achieved at the level of the regression scheme. To this end, two prominent techniques are the use of artificial neural networks and kernel ridge regression\cite{Bishop:2016ui}. We use the latter formalism here, and many further details about these techniques can be found in the rest of this volume. 
The kernel approach starts with the definition of a kernel function, which will be combined with a set of representative atomic configurations to construct the basis functions for the fit. It is a scalar function---at least when learning scalar quantities---with two input arguments, in the present case two atomic structures. Its value should quantify the similarity of the atomic configurations represented by its two arguments, and it can (but does not have to) be defined starting from their associated representations. The value should be largest when its two arguments are equal (or equivalent up to symmetry operations), and smallest for maximally different configurations. The degree to which the kernel is able to capture the variation of the function when varying the atomic configuration will determine how efficient the fit is. The better the correspondence, the fewer basis functions that are needed to achieve a given accuracy of fit.

\section{Kernel fitting}
\label{sec:krr}

We start by giving a concise account of the kernel regression fitting approach, for more details see Refs.~\onlinecite{Bishop:2016ui,Rasmussen:2006vz,Scholkopf:2002wx}. A function defined on an atomic structure is represented as a linear sum over kernel basis functions,
\begin{equation}
f (\mathcal{A})=  \sum_{\CB \in M} x_\mathcal{B} K(\mathcal{A},\mathcal{B}),
\label{eq:gapE}
\end{equation}
where the sum runs over a {\em representative set} of configurations $M$, selected from the total set $N$ of input configurations. 
The set of coefficients, combined into a vector $\bx$, are determined by solving the linear system
that is obtained when the available data (e.g. values of the target function evaluated for a set of structures) are substituted into Eq.~\eqref{eq:gapE}. In the simplest case, there is one input data value corresponding to each atomic configuration. Let $\yy$ be the vector of all available input data, and $\KK$ be the kernel matrix with rows and columns corresponding to atomic structures, so that the element of $\KK$ with row and column corresponding to structures $\CA$ and $\CB$, respectively, is $K(\CA,\CB)$. The fit is then obtained by solving a linear system in the least squares sense, i.e. minimizing the quadratic loss function,
\begin{equation}
     \loss(\bx) = \|\KK \bx - \yy\|^2 .
\end{equation}
The text book case is when the set of all configurations for which we have target data available is used in its entirety as the representative set (i.e. $N = M$), $\KK$ is square, and as long as it is invertible, the optimal solution is
\begin{equation}
    \bx = \KK^{-1}\yy.
\end{equation}
In practice, for large data sets, using all the configurations in the data set as representatives is unnecessary. In this case, $M \subset N$, the solution is given by the pseudoinverse,
\begin{equation}
    \bx_M = (\KK_{MN} \KK_{NM})^{-1}\KK_{MN} \yy_N,
\end{equation}
where we used subscripts to emphasize the set that the vector elements correspond to, e.g. $\yy \equiv \yy_N$ is the data vector with one element for each input data structure and $\bx \equiv \bx_M$ is the vector of coefficients, one for each representative configuration. The subscripts on the kernel matrix denote array slices, i.e. $\KK_{MN} = \KK_{NM}^\T$ is the rectangular matrix whose elements correspond to the kernel values between the representative configurations and the input configurations.
\nomenclature{$\CA$}{An item - structure, or atomic environment for which one wants to predict a property}
\nomenclature{$K(\mathcal{A},\mathcal{B})$}{The kernel function computed between items $\CA$ and $\CB$}
\nomenclature{$N$}{Number of input structures in the training set}
\nomenclature{$M$}{Number of structures in the representative set}
\nomenclature{$\bx$}{The vector of KRR weights, also written as $\bx_M$; the weight associated with a structure $\CB$ is indicated as $x_\CB$}
\nomenclature{$\yy$}{The vector containing the values of the target property, also written as $\yy_N$. $y_\CB$ indicates the value for the item $\CB$}
\nomenclature{$\|\cdot\|$}{The 2-norm of the quantity $\cdot$}
\nomenclature{$\|\cdot\|_F$}{The Frobenius norm of the quantity $\cdot$}
\nomenclature{$\KK$}{Kernel matrix}
\nomenclature{$\KK_{MN}$}{Slice of the kernel matrix $\KK$, corresponding to rows in set $M$ and columns in set $N$}

Using a representative set much smaller than the total number of structures has significant advantages in terms of computational cost, often with no reduction in fitting accuracy. The training cost is dominated by computing the pseudoinverse, which scales as $O(NM^2)$, which is linear in the size of the training data, $N$, and evaluating the model scales as $O(M)$, now independent of the size of the training data. These cost scalings are analogous to those of artificial neural networks with a fixed number of nodes. 

While the above solutions are formally correct, it is widely recognised that they lead to numerical instability and {\em overfitting}, i.e. they are solutions that attempt to maximise the fit to the input data, even when this might not be desirable, which is almost always the case. At first sight, this might sound surprising, since electronic structure calculations can be made deterministic, with precise convergence behaviour in terms of its parameters, such as k-point sampling, SCF tolerance, etc. However, practical calculations are never converged to machine precision, and the resulting inconsistencies between the potential energy values for different configurations is not something that is desirable to propagate to a fitted potential energy surface. The magnitude of such inconsistencies can be easily assessed before the fit is made.  Previous experience\cite{Szlachta2014,Dragoni:2018je} suggests that for large databases for materials applications using plane wave density functional theory, the error due to k-point sampling are dominant, and difficult to reduce below a few meV/atom due to the associated computational cost. 

In case we are fitting a potential energy surface with a representation that does not characterise the atomic positions of the whole system completely due to e.g. a finite cutoff, or some other choices made to gain computational efficiency, the fit is not expected to be exact, irrespective of the amount of input data. Sometimes, such {\em model error} can also be assessed {\em a priori}, e.g. in the case of a finite cutoff by measuring the contribution made to forces on an atom by other atoms beyond the cutoff.\cite{Bernstein2009,Deringer:2016uf,Fujikake:2018ce}

These two considerations suggest that allowing some  ``looseness'' in the linear system might be beneficial, because it can be exploited to allow smaller linear coefficients, making the fit more regular and thus better at extrapolation. We collect the errors we expect in the fit of each target data value on the diagonal of an $N\times N$ matrix, $\mathbf{\Lambda}$.  The common procedure to regularising  the problem is due to Tikhonov\cite{Tikhonov:2013wu}. Specifically, in ``kernel ridge regression'' (and the equivalent ``Gaussian process regression'', a Bayesian view of the same) the Tikhonov matrix is chosen to be the kernel matrix between the $M$ representative points, $\KK_{MM}$. With highly regular (``smooth'') kernel functions, this regularisation leads to smooth fits, and the sizes of the elements of $\mathbf{\Lambda}$ control the trade-off between the accuracy of the fit and smoothness. 
The corresponding solutions are 
\begin{equation}
    \bx = (\KK+\mathbf{\Lambda})^{-1}\yy.
    \label{eq:lsqsq}
\end{equation}
for the square problem, and
\begin{equation}
    \bx_M = (\KK_{MM}+\KK_{MN} \mathbf{\Lambda}^{-1}\KK_{NM})^{-1}\KK_{MN} \mathbf{\Lambda}^{-1}\yy_N,
    \label{eq:lsqrect}
\end{equation}
for the rectangular problem, where we again emphasized the index sets. This solution is equivalent to minimizing
\begin{equation}
 \| \KK\bx -\yy \|^2_{\mathbf{\Lambda}^{-1}} + \|\bx\|^2_\KK,
\end{equation}
which shows that the inverse of the tolerances in $\mathbf{\Lambda}$ are equivalent to regression weights on the different data points. \nomenclature{$\|\cdot\|_\mathbf{A}$}{The 2-norm of the quantity $\cdot$, in the metric given by $\mathbf{A}$}

With the solution of the linear system in hand, the value of the fitted function for a new structure $\CC$ can be written as 
\begin{equation}
    f(\CC) = \KK_{\CC M} \bx_M.
\end{equation}
Note that the $\KK_{\CC M}$ slice is just a vector, with elements given by the kernel between the new structure $\CC$ and the structures in the representative set $M$. 

\subsection{Selection of a representative set}\label{sub:sparse-reference}

Next we describe some ways to choose the set of representative environments over which the sum in Eq.~\eqref{eq:gapE} is taken. This can be done by simple random sampling, but we find it advantageous to use this freedom to optimise interpolation accuracy. Among the many strategies that have been proposed\cite{hartigan1979algorithm,prabhakaran2012automatic}, we discuss two that have been used successfully in the context of potential energy fitting. 
One approach to this is to maximise the dissimilarity between the elements of the representative set. A greedy algorithm to select the configurations for the representative set is ``farthest point sampling'', in which we start with a randomly selected structure, and then iteratively pick as the next structure to include the one which is farthest away from any of the structures already in the set~\cite{gonz85tcs,ceri+13jctc,bart+17sa}. The distance between two structures is measured using the ``kernel metric''\cite{Scholkopf:2002wx}, defined as
\begin{equation}
 d^2(\mathcal{A},\mathcal{B}) = K(\mathcal{A},\mathcal{A})+K(\mathcal{B},\mathcal{B})-2 K(\mathcal{A},\mathcal{B}).\label{eq:d2-kernel}
\end{equation}
This algorithm performed well for selecting molecules in regression tasks, enabling the significant reduction of the data set sizes needed to achieve a given level of accuracy\cite{de+16pccp}. 

Another technique that has been successfully used is based on matrix factorisation, which is particularly appealing when the kernel function is linear or a low order polynomial of the representation vector. Consider the matrix of feature vectors, $\DD$, in which each row is the representation vector of an input atomic configuration, such that a linear kernel is $\KK = \DD \DD^\T$. We are looking to select rows, many fewer than the total number, which span as much of the space as all rows span. This is a problem of matrix representation, specifically the representative set should serve as a low rank approximation of $\KK$ and/or  $\DD$. One solution to this is called CUR matrix decomposition\cite{maho-drin09pnas}, which can be applied to either $\KK$ or $\DD$, the latter being much cheaper if the length of the representation vectors is less than the number of data points.   

To determine the optimal set of representative configurations, we start with a singular value decomposition of $\DD$, 
\begin{equation}
\DD = USV^\T.
\end{equation}
For each data point, a {\em leverage score} is calculated, essentially the weight that the top singular vectors have on that configuration.
\begin{equation}
    \pi_\CA = \frac1k\sum_{\xi=1}^{k}(u_\CA^\xi)^2
\end{equation}
where $u_\CA^\xi$ is the element of the $\xi$-th left singular vector that corresponds to structure $\CA$. The sum runs over the first $k$ singular vectors, e.g. $k=20$ is typical. The configuration $\CA$ is included in the representative set with a probability that is proportional to its leverage score, $\pi_\CA$. 
A deterministic variant is to select one structure $\CA$ at a time -- the one with the highest leverage score -- delete the associated row from the representation matrix and orthogonalize the remaining rows of $\DD$ relative to it.
The next data point can then be selected repeating the same procedure on the now smaller feature matrix.\cite{imba+18jcp} 

Note that in the Gaussian process literature, using a subset of the data to construct the basis is called \emph{sparsification}\cite{QuinoneroCandela:2005wpb,Snelson:2005vi}, even though the approximation relies on a low rank matrix reconstruction rather than the kernel matrix being sparse. 

\subsection{Linear combination of kernels}

When fitting interatomic potentials for materials, a model is constructed for the {\em atomic energy}, sometimes called the ``site energy''. 
This is both for computational efficiency and to reduce the complexity of the functional relation between structures and properties: each atomic energy is only a function of a limited number of degrees of freedom corresponding to the coordinates of the neighbouring atoms, and can therefore be evaluated independently from any other atomic energy. In fact this is the defining characteristic of an interatomic potential, in contrast to a quantum mechanical model that explicitly includes delocalised electrons. Going from atomic energies to the total energy is trivial, the latter being the sum of the former. However, going in the other direction is not unambiguous. The total energy can be calculated from a quantum mechanical model, but the atomic energies are not defined uniquely, and it becomes part of the fitting task to find the best possible decomposition of the total energy into atomic energies. 
Treating these two transformations on the same footing helps. Suppose we want to predict the sum of function values for two (or more) configurations.  For the simple case of the sum of two energies for structures $\CA$ and $\CB$, the prediction is, trivially, just the sum of the individual function value predictions, e.g.
\begin{equation}
    E_\mathrm{tot} = E(\CA) + E(\CB) =  \KK_{\CA M}\bx_M + \KK_{\CB M}\bx_M.~\label{eq:energy-decomposition}
\end{equation}
If we define a new ``sum-kernel'' to be the sum of kernel values between a number of new configurations and the representative set,  the expression for the above total energy prediction takes the same form as the prediction of the individual function values. For some set $I$ of new configurations, let
\begin{equation}
    {}^\Sigma\KK_{M} = \sum_{\CA \in I} \KK_{\CA M},\label{eq:sum-kernel}
\end{equation}
where ${}^\Sigma\KK_{M}$ is the {\em vector} of sum-kernel values, each element of which is the sum of the kernel between all the configurations in $I$ and a given configuration in the representative set $M$.  The predicted total energy of the configurations in $I$ is then
\nomenclature{${}^\Sigma\KK$}{Sum-kernel, defined as the sum of the regular kernel over a set of configurations}
\begin{equation}
    E_\mathrm{tot} = {{}^{\Sigma}\KK_M} \bx_M.
\end{equation}
This same sum-kernel can be used to  fit the model to sum data, rather than to individual function values. This is necessary in order to fit interatomic potentials for materials systems, since only total energies, and not the atomic energies themselves,  are available from electronic structure calculations. At the same time, in order to enforce a finite short range in the interatomic potential, we must express the potential as an atomic energy. Using the sum-kernel, this is straightforward, the original functional form in Eq.~\eqref{eq:gapE} can be retained, and the we now minimise (omitting the regularisation term for clarity)
\begin{equation}
    \|{}^{\Sigma}\KK \bx - \mathbf{E}_\mathrm{tot}\|^2,
\end{equation}
where ${}^{\Sigma}\KK$ is a matrix containing the sum-kernel values for all configurations in the input database and the representative set, and the vector $\mathbf{E}_\mathrm{tot}$ is the collection of corresponding total energy data.

\subsection{Derivatives}

\newcommand{\lCnabla}{\overleftarrow{\Cnabla}}
\newcommand{\rCnabla}{\overrightarrow{\Cnabla}}
The explicit analytic functional form of Eq.~\eqref{eq:gapE} leads to analytic derivates with respect to the atomic coordinates, e.g. forces in the case of fitting an energy. 
Considering for the moment the simpler case in which we are computing the derivatives of an atom-centered quantity $f(\CA)$, we define $\nabla_\CA$ as the vector of derivatives with respect to all the atomic coordinates in structure $\CA$. We use the notation $\lCnabla$ to indicate a derivative operator that applies to the first argument of the kernel, and $\rCnabla$ to indicate a derivative that applies to the second argument.
The derivatives of $f(\CA)$ are non-zero only for atoms that belong to the structure $\CA$, and are then given by differentiating Eq.~\eqref{eq:gapE}
\begin{equation}
    \nabla_\CA f(\CA) = \sum_{\CB \in M} x_\CB \lCnabla_\CA K(\CA,\CB) = \KK_{\Cnabla\CA M} \bx_M,
\end{equation}
where we introduced the notation $\KK_{\Cnabla\CA M}$ to indicate the matrix that contains the derivatives of the kernels relative to all the relevant atomic coordinates.
Similarly to the case of sums above, the gradient of the kernel function can also be used for fitting the model not to target values, but to {\em gradient data}\cite{Solak:2003vp}. This is especially useful when the target represents a potential energy surface. When using typical electronic structure methods, the cost of computing the gradient with respect to all atomic positions is only a little bit more than the cost of computing the energy, but yields much more information, $3n$ pieces of data for an $n$-atom structure.
There are two approaches one can take to incorporate gradient information. In the first one, used in Ref.~\onlinecite{Bartok:2010fj} and subsequent work of that group\cite{bgmc2013q,Szlachta2014,bc2015q,Deringer:2016uf,jc2017q,Fujikake:2018ce,Dragoni:2018je,dpc2018q,cdklc2018,rcam2018q,Nguyen:2018kh}, the functional form for the energy is again retained to be the same as in Eq.~\eqref{eq:gapE}.
The corresponding loss function (again without regularisation) is %
\begin{equation}
    \|\KK_{\Cnabla N M} \bx_M - \by_{\Cnabla N}\|^2 ,
\end{equation}
where $\by_{\Cnabla N}$ refers to the concatenated vector of gradients on all atoms in the set of input structures and $\KK_{\Cnabla N M}$ to the corresponding matrix of kernel derivatives.  
The form of the solution for the coefficients is unchanged from Eq.~\eqref{eq:lsqsq} or Eq.~\eqref{eq:lsqrect} with $\KK_{\Cnabla NM}$ taking the role of $\KK_{NM}$. 
\nomenclature{$\lCnabla$}{Derivative operator applying to the first argument of the kernel matrix}
\nomenclature{$\rCnabla$}{Derivative operator applying to the second argument of the kernel matrix}
\nomenclature{$\KK_{ {\Cnabla \CA} \, \CB}$}{Derivative of the kernel matrix, applying to its first argument, with respect to the coordinates of atoms in structure $\CA$, with structure $\CB$ as its second argument}

In the second approach, used recently in Ref.~\onlinecite{ctspsm2017q}, derivatives of the kernel are the basis functions in the functional form of the fit,
\begin{equation}
    f(\CA) = \sum_{\CB \in M}
    \bx_{\Cnabla\CB}\cdot \rCnabla_{\CB} K(\CA,\CB),
    \label{eq:GDML}
\end{equation}
where $\bx_{\Cnabla\CB}$ contains one weight for each of the derivatives relative to the atoms in structure $\CB$.
The number of basis functions and corresponding coefficients is now much larger, $3nM$, for $n$-atom structures.  Since the model is fitted to the derivatives, given by gradients of Eq.~\eqref{eq:GDML}, the loss is%
\begin{equation}
    \|\KK_{\Cnabla N\Cnabla M} \bx_{\Cnabla M} - \by_{\Cnabla N}\|^2,
    \label{eq:gdmlloss}
\end{equation}
the target properties can be computed as
\begin{equation}
    f(\CA) = \KK_{\CA\Cnabla M} \bx_{\Cnabla M},
\end{equation}
and their derivatives as
\begin{equation}
    \nabla_{\CA} f(\CA) = \KK_{\Cnabla \CA\Cnabla M} \bx_{\Cnabla M}.
\end{equation}
The original motivation for this approach is apparent from Eq.~\eqref{eq:gdmlloss} in which the matrix can be understood as a kernel directly between atomic forces (and in case of $M=N$, between the input data forces).

Both approaches constitute valid ways of learning a function from data representing its gradients, differing only in the choice of the kernel basis. The kernel-derivative basis functions could also be used in conjunction with a reduced representative set, and it is not yet clear which approach is better, or indeed a combination: one could choose different basis functions (kernels or their derivatives) depending on the amount and kind of data available and on the size and choice of the representative set. 

\subsection{Learning from linear functionals}

We can combine the sum-kernel and the derivative kernel naturally, and write a single least squares problem for the coefficients in Eq.~\eqref{eq:gapE} that is solved to fit an interatomic potential to all available total energy, force, and virial stress data (the only condition being that the input data has to be expressible using a linear operator on function values). We define $\yy$ as the vector with $L$ components containing all the input data: all total energies, forces and virial stress components in the training database,  and $\yy'$ as the vector with $N$ components containing the \textit{unknown}  atomic energies of the $N$ atomic environments in the database,  and  $\LL$ as the linear differential operator of size $L\times N$  which connects $\yy$ with $\yy'$ such that ${\LL}\yy'=\yy$ (note that the definition of $\LL$ we use here is the transpose of that in Ref.~\onlinecite{bc2015q}). The regularised least squares problem is now to minimise
\begin{equation}
    \|\LL\KK \bx -\yy\|^2_{\mathbf{\Lambda}^{-1}} + \|\bx\|^2_\KK,
\end{equation}
and the expression for the coefficients is given by
\begin{equation}
    \bx = {\big[ \KK_{MM} + (\LL\KK_{NM})^\T  \mathbf{\Lambda}^{-1} \LL \KK_{NM}  \big]}^{-1} (\LL\KK_{NM})^\T  \mathbf{\Lambda}^{-1} \yy   \,.\label{eq:KLsol}
\end{equation}
\nomenclature{$\LL$}{Linear operator connecting the observed values $\yy$ with the unobserved atomic energies $\yy'$}

It is instructive to write down the above matrices for the simple case when the system consists of just two atoms, $A$ and $B$, with position vectors $\br_A$, $\br_B$, target total energy $E$, and target forces $\mathbf{f}_A \equiv (f_{Ax},f_{Ay},f_{Az})$ and  $\mathbf{f}_B \equiv (f_{Bx},f_{By},f_{Bz})$. The data vector is then given by 
\begin{equation}
    \yy  = [\begin{matrix}E &f_{Ax} &f_{Ay} &f_{Az} &f_{Bx} &f_{By} &f_{Bz}\end{matrix}]^\T.
\end{equation}
The aim of the fit is to determine two unknown atomic energy functions $\varepsilon_A$ and $\varepsilon_B$ as a function of the atomic environments centered around the two atoms, $\CA$ and $\CB$ respectively. The total energy is their sum, $E = \varepsilon_A + \varepsilon_B$, and the forces need to include the cross terms,
\begin{equation}
\begin{split}
    \mathbf{f}_A &= \frac{\partial \varepsilon_A}{\partial \br_A} + \frac{\partial \varepsilon_B}{\partial \br_A}, \\
    \mathbf{f}_B &= \frac{\partial \varepsilon_A}{\partial \br_B} + \frac{\partial \varepsilon_B}{\partial \br_B}.
\end{split}
\end{equation}
The representative set in this case consists of the same two atoms, so $N=M$, and the kernel matrix is square,
\begin{equation}
    \KK = \left[\begin{matrix} K(\CA,\CA) & K(\CA,\CB)\\
     K(\CB,\CA) & K(\CB,\CB)
    \end{matrix}\right],
\end{equation}
and the linear operator $\LL$ is a $7\times 2$ matrix and is given by 
\begin{equation}
    \LL = \left[\begin{matrix} 1 & 1 \\
    \lCnabla_{\br_A} & \lCnabla_{\br_{A}} \\
    \lCnabla_{\br_B} & \lCnabla_{\br_B}\\
    \end{matrix}\right],
\end{equation}
so the $\LL\KK$ matrix to be substituted into equation Eq.~\eqref{eq:KLsol} is 
\begin{widetext}
\begin{equation}
    \LL\KK = \left[\begin{matrix}
    K(\CA,\CA)+K(\CA,\CB) & K(\CB,\CA)+K(\CB,\CB) \\
    \lCnabla_{\br_A} K(\CA,\CA) + \lCnabla_{\br_A} K(\CB,\CA) & \lCnabla_{\br_A} K(\CA,\CB) + \lCnabla_{\br_A} K(\CB,\CB) \\
    \lCnabla_{\br_B} K(\CA,\CA) + \lCnabla_{\br_B} K(\CB,\CA) & \lCnabla_{\br_B} K(\CA,\CB) + \lCnabla_{\br_B} K(\CB,\CB)  \\
    \end{matrix}\right]
\end{equation}
\end{widetext}
Note that terms such as $\lCnabla_{\mathbf{r}_A} K(\CB,\CB)$ or $\lCnabla_{\mathbf{r}_A} K(\CA,\CB)$ are not zero because atom $A$ is present in the environment $\CB$ of atom $B$, and so $K(\CB,\CA)$, and also $K(\CB,\CB)$, depend on $\br_A$ explicitly.

Using the approach of Ref.~\onlinecite{ctspsm2017q} for the dimer, the kernel matrix is $6\times6$, and is given by
\begin{equation}
    \KK_{\nabla \CA\nabla \CB} = \left[
    \begin{matrix}
    \lCnabla_{\br_A}\rCnabla_{\br_A} K(\CA,\CA) & \lCnabla_{\br_A}\rCnabla_{\br_B} K(\CA,\CB) \\
    \lCnabla_{\br_B}\rCnabla_{\br_A} K(\CB,\CA) & \lCnabla_{\br_B}\rCnabla_{\br_B} K(\CB,\CB)
    \end{matrix}
    \right].
\end{equation}

In practice it is always worth using {\em all} available data, even though once the fit is converged in the limit of infinite amount of data, the information from derivatives (forces) is the same as from energies. With finite amount of data, however,  choosing the weights corresponding to energies and forces via the diagonal regulariser allows control of the fit, in the sense of its relative accuracy in reproducing energies and forces. 

\subsection{Learning multiple models simultaneously}

Being able to fit to sums of function values has an interesting consequence. It enables in a very natural way the fitting of a model that is explicitly and {\em a priori} written as a sum of terms, each using a different kernel function, perhaps even using a different representation. That this is a good idea for potential energy functions is shown by the relative success of empirical force fields both for materials and molecules, in which the total energy is written as a sum of body-ordered terms, i.e. an atomic term, a pair potential and a three-body (angle-dependent) term, etc
\begin{equation}
    E_\mathrm{tot} = \sum_i E^{(1)} + \sum_{i,j} E^{(2)}(r_{ij}) 
    + \sum_{i,j,k} E^{(3)}(r_{ij}, r_{ik}, r_{jk}) +\ldots
\end{equation}

It is notable that while pair potentials and three body potentials using various simple parametrisations are widely used in the materials modelling literature, there are few models that take advantage of the full three dimensional flexibility of the three body term. Four-body terms are almost always restricted to one-dimensional parametrisations such as a dihedral angle. The reason for this is presumably because there is little intuition about what kinds of functional forms would be appropriate---kernel fitting avoids this problem. Such a framework was introduced\cite{bc2015q} and is beginning to be used for low body order model fitting\cite{Deringer:2016uf,glie+2018,zeni+2018,Fujikake:2018ce}.  Furthermore, by bringing everything together under the kernel formalism, the above expansion can also be augmented with a many-body term which enables the systematic convergence to the true Born-Oppenheimer potential energy surface, but with the many-body term having a relatively small magnitude (because the low body order terms account for most of the energy already), which helps transferability and stability. 

The two-body term could be represented as a linear combination of kernels whose arguments are simply the interatomic distances, the three-body term is again a linear combination of kernels whose arguments are some representation of the geometry of three atoms, e.g. the one above using the three distances, but two distances and an angle are equally viable. The fit is then made to the target data of total energies and forces of atomic configurations, in complete analogy with Eq.~\eqref{eq:energy-decomposition}, and now the value of the sum-kernel is the sum of pair- and triplet-kernel values between all pairs and triplets present in the two atomic configurations. A stringent test of this scheme is that in case of a target potential energy surface that is explicitly the sum of two- and three-body terms, the fit recovers these terms explicitly from just the total energies and forces~\cite{bc2015q}. 

\section{Density-based representations and kernels}

Having summarized the algorithms that can be used to perform kernel ridge regression using atomic-scale properties and their derivatives as inputs, we now proceed to describe a framework for defining physics-based representations of local atomic environments and the kernels built from them.
In kernel ridge regression, the representations do not necessarily need to be expressed explicitly, but can also be defined implicitly by means of the kernel function $K(\mathcal{A},\mathcal{B})$, that corresponds to the scalar product of representation vectors that span a (possibly infinite-dimensional) Hilbert space~\cite{Scholkopf:2002wx}. Vectors $\ket{\mathcal{A}}$ in this ``reproducing kernel Hilbert space'' do correspond to atomic structures, and one can write formally $K(\mathcal{A},\mathcal{B})\equiv\bra{\mathcal{A}}\ket{\mathcal{B}}$ even if the kernel might be computed without ever determining the vectors explicitly. 
\nomenclature{$\ket{\CA}$}{An abstract vector that describes the input $\CA$}
\nomenclature{$\bra{\CA}\ket{\CB}$}{The scalar product between the features associated with $\CA$ and $\CB$. Could be either an explicit scalar product, or an abstract notation equivalent to $K(\CA,\CB)$}

The reader trained in quantum mechanics will recognize an isomorphism between representations and the state vectors on one hand, and kernels and expectation values on the other. This analogy suggests that it may be beneficial to formulate atomic-scale representations using a formalism that mimics Dirac notation.
Whereas in a quantum mechanical setting the physical symmetries of the problem are built into the Hamiltonian, in a machine-learning setting they are more conveniently included in the representation itself, that should be made invariant to basic symmetries such as atom labelling, rigid translations and rotations.
In this section we show how starting from these intuitions one can build a very abstract description of a molecular structure that is naturally invariant with respect to the physical symmetries, based on a representation of the atom density. 

Translational and rotational symmetries can be included by decomposing the structure into a collection of local environments, and by explicit symmetrization over the $SO(3)$ group.
In fact, it has been recently shown~\cite{density-arxiv} how this construction leads naturally to the SOAP representation and kernel \cite{bart+13prb}, and to several other popular choices of density-based representations -- from Behler-Parrinello symmetry functions \cite{behl-parr07prl}, to voxel density representations \cite{kaji+17sr} to the binning of the pair correlation function \cite{schu+14prb} -- that can be regarded as different projections of the same smooth atomic amplitude.
A peculiarity of the SOAP framework is that it is formulated very naturally in terms of a kernel, that corresponds to the symmetrized overlap of atomic densities, and that it allows one to explicitly compute the representations whose scalar product constitutes the kernel function, which allows one to go back and forth between a kernel and a representation language. 
The atomic environmental representations can then be modified to generate non-linear kernels, as well as combined into global structural kernels. We will briefly discuss different possible approaches to the latter, either by simple linear combination of the local representations, or by a more sophisticated procedure that takes into account the most effective matching between pairs of environments in the two structures that are being compared.

\subsection{A Dirac notation for structural representations}

Let us introduce an abstract notation to describe atomistic structures in terms of the positions and chemical nature of the atoms that compose them~\cite{density-arxiv}. 
Taking inspiration from Dirac notation for quantum mechanical states, we associate a ket $\ket{\mathcal{A}}$ with each configuration. 
Let us start with a simple example to see how such a formalism can be introduced and used. Much like in the case of quantum states, 
we can define a concrete representation of the ket associated with a structure in terms of positions and chemical species, e.g.
\begin{equation}
    \bra{\br}\ket{\mathcal{A}} = \sum_i g_i(\br -\br_i) \ket{\alpha_i},
    \label{eq:dirac-A}
\end{equation}
where the position of each atom is represented by a smooth density $g_i$ 
(that in principle could depend on the nuclear charge and the position of atom $i$) 
and the kets $\ket{\alpha_i}$ contain the information on the  nuclear charge of each atom. 
\nomenclature{$g_i(\br)$}{A smooth function - typically a Gaussian that is used to represent the density associated with atom $i$}
\nomenclature{$\ket{\alpha}$}{An abstract vector that represents the chemical species $\alpha$}
The Dirac notation lends itself naturally to the definition of overlap kernels between structures, 
$\bra{\mathcal{A}}\ket{\mathcal{B}} $. To compute such an integral, one can use the position representation
and assume that the kets associated with different elements are orthonormal:
\begin{align}
\begin{split}
\bra{\mathcal{A}}\ket{\mathcal{B}} =& \int \dbr \, \bra{\mathcal{A}}\ket{\br} \bra{\br}\ket{\mathcal{B}}\\
=&
\sum_{ij}
\int \dbr \, g_i^A(\br -\br_i^A)^\star
g_j^B(\br -\br_j^B)\bra{\alpha_i^A}\ket{\alpha_j^B}\\
=& \sum_\alpha\sum_{i,j\in\{\alpha\}} \int \dbr \, g_i^A(\br -\br_i^A)^\star
g_j^B(\br -\br_j^B).
\end{split}
    \label{eq:dirac-AB}
\end{align}
\nomenclature{$\br$}{Position in 3D Cartesian coordinates.}
\nomenclature{$r$}{The modulus of the vector $\br$}
\nomenclature{$\hat{\br}$}{The unit vector $\br/r$}
\nomenclature{$\br_i$}{Position of the $i$-th atom}
\nomenclature{$\br_{ij}$}{Displacement vector $\br_i-\br_j$ between the $i$-th and $j$-th atoms}

This density-based representation would not be in itself very useful, as the kernel is not invariant to relative rotations of the structures, and not even to the absolute position of the two structures in space, or their periodic representation. Nevertheless, it can be taken as the starting point to introduce many of the most successful feature representations that have been used in recent years for machine-learning of materials and molecules. 

To see how, one can take inspiration from linear-scaling electronic structure methods, and the nearsightedness principle for electronic matter~\cite{yang91prl,gall-parr92prl,goed99rmp,prod-kohn05pnas}.
We then shift the attention from the description of complete structures to that of spherical atomic environments, that one can conveniently center on top of each atom.
An atom-centered representation arises naturally from the symmetrization over the translation group of tensor products of the representation Eq.~\eqref{eq:dirac-A}~\cite{density-arxiv}, and is also consistent with the atom-centered potentials that have been discussed in the previous Section as an obvious application of this framework. 

We will use the notation $\ket{\xj}$ to indicate an environment centered around the $j$-th atom in a structure, and express it in the position representation as 
\begin{align}
    \bra{\br}\ket{\xj} = \sum_i f_c(r_{ij})
    g_{ij}(\br-\br_{ij})\ket{\alpha_i}
    \label{eq:dirac-xj}
\end{align}
where $f_c(r_{ij})$ is a cutoff function that restricts the environment to a spherical region centered on the atom, for the sake of computational efficiency and/or localization of the density information. The atom-centered smoothing functions are typically taken to be uniform-width Gaussians, but it would be easy to generalize the expression to include a dependency on the atomic species and/or the distance of an atom from the center of the environment, which could be used to e.g. reduce the resolution of the representation at the periphery of the environment, or adapt the smoothing length scale to each  atomic species. 

Note that one could also combine the density contributions from atoms of the same species into a species-dependent atomic amplitude,
\begin{equation}
\bra{\alpha\br}\ket{\xj}=\psi^\alpha_{\xj}(\br) = \sum_{i\in \alpha} f_c(r_{ij})
    g_{ij}(\br-\br_{ij}),
\end{equation}
and then write
\begin{equation}
    \bra{\br}\ket{\xj} = \sum_\alpha \psi^\alpha_{\xj}(\br)\ket{\alpha}.
    \label{eq:dirac-xjalpha}
\nomenclature{$\psi_{\xj}^\alpha(\br)$}{The atom density of species $\alpha$ centered around the $j$-th atom}
\end{equation}

This notation is very useful to reveal how different representations can be seen as alternative representations of the same abstract ket. For instance, one can expand the atom density in orthogonal radial functions $R_n(r)$ and spherical harmonics. The coefficients in such an expansion can be written as 
\begin{equation}
\begin{split}
\bra{\alpha nlm}\ket{\xj} = & \int \dbr \bra{nlm}\ket{\br} \bra{\alpha\br}\ket{\xj} \\
=&\int \id r \id\hat{\br}\, r^2 R_n(r) Y^l_m(\hat{\br}) \psi_\xj^\alpha (r\hat{\br}).
\end{split}
    \label{eq:dirac-nlm}
\end{equation}
\nomenclature{$Y^l_m(\hat{\br})$}{The $l,m$-th spherical harmonic}
\nomenclature{$R_n(r)$}{The $n$-th orthogonal radial basis function}

As another example, Behler-Parrinello atom-centered symmetry functions, that have been used in the construction of artificial neural network based interatomic potentials for materials~\cite{behl-parr07prl,eshe+12prl,mora+16pnas,chen+16jpcl} and molecules~\cite{smit+17cs} can be written by setting the basis functions to be delta distributions $g_{ij}(\br-\br_{ij})=\delta(\br-\br_{ij})$, and averaging the atom density with an appropriate pair  weighting function $G_2$, e.g.
\begin{equation}
\begin{split}
\bra{\alpha\beta G_2}\ket{\xj} =& 
\bra{\alpha}\ket{\alpha_j}
\int \dbr\, G_2(r) \bra{\beta\br}\ket{\xj} \\
=& \, \delta_{\alpha_j\alpha}\sum_{i\in\{\beta\}} f_c(r_{ij}) G_2(r_{ij})
\end{split}
\label{eq:dirac-bpsf}
\end{equation}
The basis functions of the Spectral Neighbour Analysis Potential\cite{Thompson:2015dw} also start with the same density, and expands it in hyperspherical harmonics as introduced in Ref.~\onlinecite{Bartok:2010fj}. 

\subsection{Smooth Overlap of Atomic Positions}

It is clear that a density-based representation such as Eq.~\eqref{eq:dirac-xj} is invariant to translations of the entire structure, but not to rotations that would change the orientation of the atomic neighbour amplitude. This reflects the fact that scalar products of the form $\bra{\xj}\ket{\xk}$ depend on the relative orientation of the environments being compared.
In the Smooth Overlap of Atomic Positions (SOAP) framework, we define a symmetrized version of the overlap kernel, using the Haar integral\cite{Haar:1933ic} of the rotation group,
\begin{align}
    K^{(\nu)}(\xj,\xk)=& 
    \int \drhat \left| \bra{\xj}\hat{R}\ket{\xk} \right|^{\nu} = \bra{\xj^{(\nu)}}\ket{\xk^{(\nu)}}
    \label{eq:soap-integral}
    \nomenclature{$\ket{\xj^{(\nu)}}$}{The spherically-averaged SOAP representation of order $\nu$}
\end{align}
where the integral is performed over all possible rotation matrices. 
If the base kernel is raised to the $\nu$-th power, the average preserves information on the correlations between atoms up to the $(\nu+1)$-th order~\cite{glie+2018}.
As we will show below, a crucial feature of the SOAP framework is that an explicit expression for the symmetrized representation vectors $\ket{\xj^{(\nu)}}$ can be given. In fact, an alternative derivation of the SOAP framework can be achieved by symmetrizing directly tensor products of the translationally invariant ket Eq.~\eqref{eq:dirac-nlm}~\cite{density-arxiv}. 

The complexity of the SOAP features is quite manageable for $\nu=1,2$, but becomes increasingly cumbersome for higher $\nu$. 
An effective description of higher-order interactions, that does not increase too much the complexity of the analytical evaluation of Eq.~\eqref{eq:soap-integral}, can be obtained by manipulating the $\nu=2$ kernel, e.g. by taking a non-linear function of it. In practice it has been found that raising it to a power $\zeta$, and normalizing it to one
\begin{equation}
\bra{\xj^{(2)}}\ket{\xk^{(2)}}_\zeta =
    \frac {
    \bra{\xj^{(2)}}\ket{\xk^{(2)}}^\zeta }{
    \sqrt{
    \bra{\xj^{(2)}}\ket{\xj^{(2)}}^\zeta
    \bra{\xk^{(2)}}\ket{\xk^{(2)}}^\zeta
    }
    }
    \label{eq:soap-kernel-zeta}
\end{equation}
is sufficient to include many-body contributions in the final kernel. 
\nomenclature{$\bra{\xj^{(\nu)}}\ket{\xk^{(\nu)}}_\zeta$}{The normalized SOAP kernel of order $\nu$ and non-linear exponent $\zeta$}

Using the Dirac notation, it is easy to see how one can give an explicit representation of the $SO(3)$ symmetrized ket for the case with $\nu=1,2$. 
Using a spherical harmonics expansion of $\ket{\xj}$ it is very natural to perform the rotational average analytically by introducing the Wigner matrix associated with the rotation, $\bra{l m}\hat{R}\ket{l' m'} = \delta_{ll'}D^l_{mm'}(\hat{R})$ 
\begin{equation}
\begin{split}
\int \drhat \sum_{\alpha n l m} \bra{\xj}\ket{\alpha n l m} \bra{\alpha n l m}\hat{R}\ket{\xk}=\\
\sum_{\alpha n l m m'} \bra{\xj}\ket{\alpha n l m}
\bra{\alpha n l m'}\ket{\xk}
\int \drhat\, D^l_{mm'}(\hat{R}) 
\end{split}
\label{eq:soap-integral-1}
\nomenclature{$D^l_{mm'}(\hat{R})$}{The Wigner rotation matrix associated with the rotation $\hat{R}$}
\end{equation}
which simplifies greatly due to the properties of the Wigner matrices. Only the term with $l=0$ survives, which makes it possible to write explicitly the $\nu=1$ symmetrized SOAP representations in terms of the spherical harmonics coefficients
\begin{equation}
\bra{\alpha n}\ket{\xk^{(1)}} =\sqrt{8\pi^2}\bra{\alpha n 0 0}\ket{\xk},
\end{equation}
which corresponds to the simple kernel
\begin{equation}
\bra{\xj^{(1)}}\ket{\xk^{(1)}}=\sum_{\alpha n} \bra{\xj^{(1)}}\ket{\alpha n} \bra{\alpha n}\ket{\xk^{(1)}}.
\end{equation}
A position representation of the $\nu=1$ representation $\bra{r}\ket{\xk^{(1)}}$ yields naturally the rotational average of $\bra{\br}\ket{\xk}$.
This can be seen by expressing $K^{(1)}(\xj,\xk)$ in a position basis
\begin{equation}
\begin{split}
\bra{\alpha\xj^{(1)}}\ket{\alpha\xk^{(1)}} = &
\int \drhat \int \id \br\,
\psi^\alpha_\xj(\br) \psi^\alpha_\xk(\hat{R}\br) \\
= & 
32\pi^3 \int \id r\, r^2 \bar{\psi}^\alpha_\xj(r) \bar{\psi}^\alpha_\xk(r)
\end{split}
\end{equation}
where we have defined the rotationally-averaged atom density
\begin{equation}
\bar{\psi}^\alpha_\xj(r) =\frac{1}{4\pi}\int \id \hat{\br}\, \psi^\alpha_\xj(r \hat{\br}) = \frac{1}{\sqrt{32\pi r^3}}\bra{\alpha r}\ket{\xj^{(1)}},\label{eq:xi1-r}
\end{equation}
which is thus closely related to the pair correlation function around the tagged atom.
Similar representations have been used for machine-learning of molecules and materials~\cite{schu+14prb,fabe+15ijqc}, revealing once more the intimate relationships between different atom-density based representations. 

The $\nu=1$ representation integrates away all angular correlations and therefore does not provide a unique representation of an environment. 
The representations with $\nu=2$ provide information on 3-body correlations, and can also be obtained relatively easily in closed form. The Haar integral now contains the product of two Wigner matrices. Exploiting their orthogonality relations, one obtains
\begin{equation}
\begin{split}
\int \drhat \left| \sum_{\alpha n l m} \bra{\xj}\ket{\alpha n l m} \bra{\alpha n l m}\hat{R}\ket{\xk} \right|^2=\\
\sum_{\alpha n \alpha' n' l}
\bra{\xj^{(2)}}\ket{\alpha n \alpha' n' l} \bra{\alpha n\alpha' n' l}\ket{\xk^{(2)}}
\end{split}
\label{eq:soap2-integral}
\end{equation}
where the $\nu=2$ symmetrized SOAP representations read
\begin{align}
\bra{\alpha n \alpha' n'l}\ket{\xj^{(2)}} =&\sqrt{\frac{8\pi^2}{2l+1}}\sum_{m} \bra{\xj}\ket{\alpha nlm}
\bra{\alpha' n'lm}\ket{\xj}.
\nomenclature{$\bra{\alpha n \alpha' n'l}\ket{\xj^{(2)}}$}{The radial/spherical representation of the SOAP $\nu=2$ vector, corresponding to the power spectrum between species $\alpha$ and $\alpha'$}
\label{eq:soap-representation}
\end{align}
This notation corresponds to the power-spectrum components introduced in Refs.~\cite{bart+13prb,de+16pccp},
$\bra{\alpha n \alpha' n'l}\ket{\xj^{(2)}}\equiv p^{\alpha\alpha'}_{nn'l}(\xj)$.
Note also that, while the representation of the symmetrized kets in terms of the $nlm$ expansion is very convenient,
it is not the only possibility.
Similar to Eq.~\eqref{eq:xi1-r}, an explicit position representation can be obtained for $\bra{\alpha\br_1\alpha'\br_2}\ket{\xk^{(2)}}$,  that provides a complete representation of the 3-body rotationally-invariant correlations. The 3-body symmetry functions of the Behler-Parrinello kind can be seen as projections of this representation, similarly to the case of 2-body functions in Eq.~\eqref{eq:dirac-bpsf}. 

The case of $\nu=3$ leads to an explicit representation of the ket that is proportional to the bispectrum of the environment~\cite{bart+13prb}
\begin{equation}
\begin{split}
&\bra{\alpha_1 n_1 l_1 \alpha_2 n_2 l_2 \alpha n l}\ket{\xj^{(3)}} \propto \sum_{m m_1 m_2} \bra{\xj}\ket{\alpha n l m} \\
\times&\bra{\alpha_1 n_1 l_1 m_1}\ket{\xj}
\bra{\alpha_2 n_2 l_2 m_2}\ket{\xj} \bra{l_1\, m_1\, l_2\, m_2}\ket{l\, m}.
\end{split}
\label{eq:bispectrum}
\nomenclature{ $\bra{l_1\, m_1\, l_2\, m_2}\ket{l\, m}$}{A Clebsch-Gordan coefficient}
\end{equation}
While the dimensionality of this representation makes it impractical unless somehow sparsified, it does give direct access to higher-order correlations. An interesting detail is that $\ket{\xj^{(3)}}$, contrary to the $\nu=1,2$ cases, is not invariant to mirror symmetry, which makes it capable of distinguishing enantiomers.

Finally, one should note that the normalization of the kernel Eq.~\eqref{eq:soap-kernel-zeta} can be achieved by normalizing the SOAP vector, so that an explicit representation of the normalized feature vector is possible. While in principle one could write out an explicit representation that yields the kernel for $\zeta>1$, it would contain an exponentially increasing number of terms. As in the case of  $\ket{\xj^{(3)}}$, this only makes sense if combined with a sparsification procedure.

\subsection{Kernel operators and feature optimization}

Provided one takes a long-range environmental cutoff, and chooses a kernel that can represent high orders of many-body interactions, a density-based representation of atomic structures should provide a complete description of any atomic structure and -- given a sufficiently complete training set -- predict any atomistic property with arbitrary accuracy. 
In practice, obviously, the accuracy of a model depends on the details of the representation, which is why different representations or kernels provide different levels of accuracy for the same training and test set~\cite{fabe+17jctc}. The performance of a set of representations can be improved by modifying them so that they represent more efficiently the relations between structure and properties. 

This kind of optimizations are best understood in terms of changes to the translationally-invariant environmental ket Eq.~\eqref{eq:dirac-xj}, and can be described, in an abstract and basis-set independent manner as a Hermitian operator acting on the ket, 
\begin{equation}
\ket{\xj} \rightarrow \hat{U} \ket{\xj}.
\end{equation}
The most general form of this operator that makes it rotationally-invariant -- so that it commutes with the rotation matrix in the definition of the SOAP kernel Eq.~\eqref{eq:soap-integral} -- is readily expressed in the  $\{\ket{\alpha nlm}\}$ basis~\cite{density-arxiv}:
\begin{equation}
    \bra{\alpha nlm}\hat{U}\ket{\alpha' n'l'm'} = \, \delta_{ll'} \delta_{mm'}  \bra{\alpha
    n l}\hat{U}\ket{\alpha' n' l'}.
\label{eq:wop-nalpha}    
\end{equation}

While this is the most general form of the operator that is consistent with $SO(3)$ symmetry, one can use simpler forms to represent feature space transformations that can be easily understood. 
For instance, taking
\begin{equation}
\bra{\alpha n l}\hat{U}\ket{\alpha' n' l'} = u_n \delta_{\alpha\alpha'} \delta_{nn'} \delta_{ll'}
\end{equation}
corresponds to a scaling of the smooth atom density according to the distance from the center. 
This kind of scaling has been shown to improve significantly the performance of SOAP~\cite{alchemy-arxiv}, as well as other density-based representations~\cite{Huang2016,fabe+18jcp}.

Another simple form of the transformation matrix involves only the ``chemical'' channels
\begin{equation}
\bra{\alpha n l}\hat{U}\ket{\alpha' n' l'} = u_{\alpha\alpha'} \delta_{nn'} \delta_{ll'}.
\end{equation}
This operator amounts at a change of representation for the elemental space. It is easy to see that $\bra{\alpha}\hat{U}^\dagger \hat{U}\ket{\alpha'}= \kappa_{\alpha\alpha'}$ is nothing but the ``alchemical similarity matrix'' between elements that has been shown to improve the accuracy of SOAP in the presence of multiple atomic species~\cite{de+16pccp,bart+17sa}. 
What is more, by writing a low-rank approximation of $\hat{U}\approx \sum_{J\alpha} u_J\alpha \ket{J}\bra{\alpha}$ one can express atomic density in terms of a small number of ``chemical archetypes'', improving dramatically both the accuracy and the computational cost of machine-learning models that involve more than a handful of elements~\cite{alchemy-arxiv}. 
Note that this transformation can be applied at the level of the translationally-invariant representation, where one can write
\begin{equation}
 \psi^J_{\xj}(\br) = \bra{J\br}\ket{\xj} = \sum_\alpha u_{J\alpha} \bra{\alpha \br}\ket{\xj}
\end{equation}
that makes it evident how the action of this particular $\hat{U}$ operator amounts at using linear combination of atomic densities in which each specie is given weights that can be optimized by cross-validation.

\subsection{$\lambda$-SOAP: Symmetry-Adapted Gaussian Process Regression}

When building a machine-learning model for a tensorial property $\mathbf{T}$, one should consider that the target is not invariant under the action of a symmetry operation (e.g. a rotation) but transforms covariantly. 
The most effective strategy to encode the appropriate covariance properties in the model involves the decomposition of the tensor into its irreducible spherical components, i.e. combinations of the elements of the tensor that transform as the spherical harmonics of order $\lambda$~\cite{Varshalovich:1988ul}. For these irreducible components,
\begin{equation}
    T_{\lambda\mu}(\hat{R}\xj) = \sum_{\mu'} D^\lambda_{\mu\mu'}(\hat{R}) T_{\lambda\mu'}(\xj)
\nomenclature{$T_{\lambda\mu}$}{The $\mu$-th component of the irreducible spherical component of order $\lambda$ for the tensorial quantity $\mathbf{T}$}
\end{equation}
As shown in Ref.~\cite{glie+17prb} for the case of vectors and in Ref.~\cite{gris+18prl} for tensors of arbitrary order, one has to consider a matrix-valued kernel that describes the geometric relationship between the different components of $\mathbf{T}_\lambda$, which can be obtained by including an additional Wigner matrix $D^\lambda_{\mu\mu'}(\hat{R})$  in the Haar integral
\begin{equation}
\bra{\calx_{j,\lambda\mu}^{(\nu)}}\ket{\calx_{k,\lambda\mu'}^{(\nu)}} = 
    \int \drhat\, D^\lambda_{\mu\mu'}(\hat{R}) \left| \bra{\xj}\hat{R}\ket{\xk} \right|^\nu.
    \label{eq:l-soap-integral}
\end{equation}
For the case with $\nu=2$ the symmetrized kets can be written explicitly based on a $\alpha nlm$ expansion of the atom density
\begin{equation}
\begin{split}
\bra{\alpha n l \alpha' n'l'}\ket{\calx_{j,\lambda\mu}^{(2)}}  =&\sqrt{\frac{8\pi^2}{2l+1}} \sum_{mm'} \bra{\xj}\ket{\alpha nlm} \\
 \times & \bra{\alpha' n'l'm'}\ket{\xj} \bra{l\, m\, l'\,\, {-m'}}\ket{\lambda\,\, {-\mu}}%
\label{eq:l-soap-representation}
\end{split}
\nomenclature{$\ket{\calx_{j,\lambda\mu}^{(\nu)}}$}{The $\lambda$-SOAP representation of order $\nu$, corresponding to the irreducible spherical component $\lambda\mu$ centered on atom $j$}
\end{equation}
We write Eq.~\eqref{eq:l-soap-representation} in this form because it is somewhat symmetric, but the properties of the CG coefficients require that $m'=m+\mu$ so the expression can be evaluated with a single sum. 
Furthermore, the expression evaluates to zero whenever $\left|l-l'\right|<\lambda$, which reduces the number of elements that 
must be evaluated and stored, and makes it clear that Eq.~\eqref{eq:l-soap-representation}
reduces to the scalar SOAP Eq.~\eqref{eq:soap-representation} when $\lambda=0$.

When using a linear model, each of the the symmetry-adapted representations Eq.~\eqref{eq:l-soap-representation} can be used to represent tensorial components that transform as $Y^\lambda_\mu$. 
Linearity, in this case, is necessary for preserving the symmetry properties of the $\lambda$-SOAP~\cite{alpha-arxiv}. 
A non-linear model, however, can be obtained by scaling each $\bra{\alpha n l \alpha'n'l'}\ket{\calx_{j,\lambda\mu}^{(2)}}$ by a (in principle different) non-linear function of some $\lambda=0$ representations.
In the kernel language, a high-order version of the $\lambda$-SOAP kernel can be introduced with an expression analogous to Eq.~\eqref{eq:soap-kernel-zeta}:
\begin{equation}
\bra{\calx_{j,\lambda\mu}^{(2)}}\ket{\calx_{k,\lambda\mu'}^{(2)}}_\zeta =  
\frac {\bra{\calx_{j,\lambda\mu}^{(2)}}\ket{\calx_{k,\lambda\mu'}^{(2)}}
    \bra{\xj^{(2)}}\ket{\xk^{(2)}}_{\zeta-1} }{
\left\|\bra{\calx_{j,\lambda\boldsymbol{\mu}}^{(2)}}\ket{\calx_{j,\lambda\boldsymbol{\mu}^\T}^{(2)}}\right\|_F \left\|\bra{\calx_{k,\lambda\boldsymbol{\mu}}^{(2)}}\ket{\calx_{k,\lambda\boldsymbol{\mu}^\T}^{(2)}}\right\|_F
    },
    \label{eq:l-soap-kernel-zeta}
\end{equation}
where $\left\|\cdot\right\|_F$ indicates the Frobenius norm and $\bra{\xj^{(2)}}\ket{\xk^{(2)}}_{\zeta-1}$ is a (scalar) SOAP kernel. This second term makes the overall kernel non-linear, without affecting the symmetry properties of the overall tensorial kernel.

\subsection{Computing SOAP representations efficiently}

A practical calculation of both scalar and tensorial $\nu=2$ SOAP representations $\bra{\alpha n l \alpha' n' l'}\ket{\calx_{j,\lambda\mu}^{(2)}}$ requires the evaluation of the expansion coefficients $\bra{\alpha n l m}\ket{\xj}$. 
Let us start with the atom density written in the position representation, according to Eq.~\eqref{eq:dirac-nlm}, 
and consider the case in which $\psi^{\alpha}_{\mathcal{X}}(\mathbf r)$ is written as a superposition 
of spherical Gaussian functions of width $\sigma$ placed at the positions of the atoms of type $\alpha$.
Then, the spherical harmonics projection in Eq.~\eqref{eq:dirac-nlm} can be carried out analytically, leading to:
\begin{equation}
\begin{split}\label{semiformal_coeffs}
    \bra{\alpha n l m}\ket{\xj} = &\sum_{i \in \alpha} Y_{lm}(\hat{ \br}_{ij})\ e^{-\frac{r^2_{ij}}{2\sigma^2}}\ \times \\& \times \int_0^\infty \mathrm{d}r\ r^2\ R_n(r)
e^{-\frac{r^2}{2\sigma^2}} \iota_l\left(\frac{r r_{ij}}{\sigma^2}\right) \
\end{split}
\end{equation}
where the sum runs over all neighbouring atoms of type $\alpha$ and $\iota_l$ indicates a modified spherical Bessel function of the first kind. 
It is convenient to choose a form for the orthogonal radial basis functions $R_{n}(r)$ that makes it possible to perform the radial integration analytically.

One possible choice starts by using Gaussian type orbitals as non-orthogonal primitive functions $\tilde{R}_k(r)$
\begin{equation}
    \tilde{R}_{k}(r) = \mathcal{N}_k\ r^{k} \exp{-\frac{1}{2}\left(\frac{r}{\sigma_{k}}\right)^2}, 
\end{equation}
where $\mathcal{N}_k$ is a normalization factor, such that $\int_0^\infty dr r^2 \tilde{R}^2_{k}(r) = 1$. The set of Gaussian widths $\{\sigma_k\}$ can be chosen to span effectively the radial interval involved in the environment definition. 
Assuming that the smooth cutoff function approaches one at a distance $r_\text{cut}-\delta r_\text{cut}$, one could take $\sigma_k = (r_\text{cut}-\delta r_\text{cut}) \max(\sqrt{k},1)/n_\text{max}$, 
that gives functions that are peaked at equally-spaced positions in the range between 0 and $r_\text{cut}-\delta r_\text{cut}$.

While the $\tilde{R}_k(r)$ are not themselves orthogonal, they can be used to write orthogonal basis functions $R_n(r)=\sum_k S^{-1/2}_{nk} \tilde{R}_k(r) $, where the overlap matrix $S_{kk'} = \int \mathrm{d}r r^2 \tilde{R}_k(r) \tilde{R}_{k'}(r) $ can be computed analytically.
The full decomposition of the translationally-invariant environmental ket can then be obtained without recourse to numerical integration.

Once the spherical decomposition of the atomic density has been obtained, the coefficients can be combined to give the SOAP representations of order 1 and 2. 
Particularly in the presence of many different chemical species, the number of components can become enormous. Ignoring for simplicity a few symmetries, and the fact that if all species do not appear in every environment it is possible to store a sparse representation nof the representation, the power spectrum contains a number of components of the order of $n_\text{species}^2 n_\text{max}^2  l_\text{max} $, which can easily reach into the tens of thousands. 
In the case of the tensorial $\lambda$-SOAP the number increases further to $ \lambda^2 n_\text{species}^2 n_\text{max}^2  l_\text{max}$.
It is however not necessary to compute and store all of these representations: each of them, or any linear combination, is a spherical invariant (covariant) description of the environment and can be used separately as a representation. 
This can be exploited to reduce dramatically the computational cost and the memory footprint of a SOAP calculation, determining a low-rank approximation of the representation. 
One can use dimensionality reduction techniques similar to those discussed in Section~\ref{sub:sparse-reference} to identify the most suitable reference structures. As shown in Ref.~\citenum{imba+18jcp}, both CUR decomposition and a greedy selection strategy based on farthest point sampling make it possible to reduce by more than 95\%{} the number of SOAP representations that are needed to predict the energy of small organic molecules with chemical accuracy.

\subsection{Back to the structures}

Whenever one is interested in computing properties that are associated to individual atoms (for instance their NMR chemical shieldings, or the forces) one can use directly the representations corresponding to each environment, or the kernel between two environments, as the basis for a linear or non-linear regression model.
As discussed in Section~\ref{sec:krr}, it is often the case that one is interested in using as structure labels some properties that are instead associated with the entirety of a structure -- e.g. its cohesive energy, its dielectric constant, etc.
In these cases a ridge regression model should be used that is based on ``global'' kernels between the structures, $K(\mathcal{A},\mathcal{B})$, rather than those between individual atom-centered environments.
This is reflected in how the kernels between environments should be combined to give a kernel that is suitable to represent the relation between local environments and the overall property of a structure. 
When the target property can be seen as an additive combination of local, atom-centered contributions, the most natural (and straightforward) choice, that is consistent with Eq.~\eqref{eq:sum-kernel}, is
\begin{equation}
K(\CA,\CB) = \sum_{j\in\CA,k\in\CB}K(\xj,\xk).
~\label{eq:gkernel-sum}
\end{equation}
It is worth stressing that in the case where the environment kernel is a linear kernel based on SOAP representations, this sum-kernel can be written in terms of a global representation associated with the entire structure,
\begin{equation}
K(\CA,\CB) =\bra{\mathcal{A}^{(\nu)}}\ket{\mathcal{B}^{(\nu)}}, \label{eq:gkernel-dirac}
\end{equation}
where we introduced 
\begin{equation}
\ket{\mathcal{A}^{(\nu)}} = \sum_{j\in \mathcal{A}} \ket{\xj^{(\nu)}}. \label{eq:gkernel-ket}
\end{equation}
An alternative way to combine the information from individual environments in a symmetrized global kernel corresponds to averaging the Fourier coefficients of each environment,
\begin{equation}
\bra{\alpha nlm}\ket{\CA}=\sum_{j\in\CA}\bra{\alpha nlm}\ket{\xj}
\end{equation}
and then taking the Haar integral of the resulting sum. For instance, for $\nu=2$, 
\begin{equation}
\bra{\alpha n \alpha' n'l}\ket{\bar{\CA}^{(2)}}=\sum_{m}\bra{\alpha nlm}\ket{\CA} \bra{\CA}\ket{\alpha' n'lm}.
\end{equation}

The form Eq.~\eqref{eq:gkernel-sum} is more general, and one can readily introduce non-linear kernels such as $\bra{\xj^{(\nu)}}\ket{\xk^{(\nu)}}_\zeta$ for which an explicit expression for the representations would be too cumbersome. 
Eq.~\eqref{eq:gkernel-sum} also suggests that the combination of environment kernels could be generalized by introducing a weighting matrix
\begin{equation}
K_W(\CA,\CB) = \sum_{j\in \mathcal{A}, k\in \mathcal{B}} W_{jk}(\mathcal{A},\mathcal{B}) K(\xj,\xk). \label{eq:gkernel-weights}
\end{equation}
One could for instance determine the importance of each environment within a structure, and set $W_{jk}(\mathcal{A},\mathcal{B})=w_j(\mathcal{A}) w_k(\mathcal{B})$. 
Alternatively, one can use techniques from optimal transport theory~\cite{cutu13nips} to define an entropy-regularized matching (REMatch) procedure~\cite{de+16pccp}, in which $W_{jk}$ is a doubly stochastic matrix that matches the most similar environments in the two structures, disregarding the environmental kernels between very dissimilar environments
\begin{equation}
\begin{split}
\mathbf{W}(\mathcal{A},\mathcal{B})&=\operatorname*{argmin}_{\mathbf{W} \in \mathcal{U}(N_\mathcal{A},N_\mathcal{B})} \sum_{jk} W_{jk} \left[d^2(\xj,\xk) + \gamma \ln W_{jk} \right],
\end{split}
\end{equation}
where $d^2$ indicates the kernel-induced squared distance Eq.~\eqref{eq:d2-kernel}.
The parameter $\gamma$ weights the entropy regularization and makes it possible to interpolate between strict matching of the most similar pairs of environments ($\gamma\rightarrow 0$) to an average kernel that weights all pairs equally ($\gamma\rightarrow\infty$). 
Although this construction complicates considerably the combination of local kernels, it provides a strategy to introduce an element of non-locality in the comparison between structures.
Given the cost of computing the REMatch kernel, and the fact that it prevents using some sparsification strategies that act at the level of individual environments, this method should be used when the target property is expected to exhibit very strong non-additive behavior, e.g. when just one portion of the system is involved -- for instance when determining the activity of a drug molecule, a problem for which REMatch has been shown to improve dramatically the accuracy of the ML model~\cite{bart+17sa}.

\subsection{Multi-kernel learning}

We have shown that SOAP representations can be seen as just one possible embodiement of a general class of rotationally-symmetrized density-based representations, that also encompasses other popular representations for atomic-scale machine learning, and that can be tuned to a great extent, e.g. by changing the way different components are weighted.
The fact that different representations can be computed within the same formalism does not imply they are fully equivalent: each expression or kernel emphasizes different components of the structure/property relations. For instance, kernels with varying radial scaling or cutoff distance focus the machine-learning model on short, mid or long-range interactions. 
It is then natural to consider whether a better overall model can be constructed by combining representations that are associated with different cutoff distances, or different levels of body order expansions. This can be achieved by a weighted combination of kernels of the form
\begin{equation}
K_\text{tot}(\CA,\CB) = \sum_\aleph w_\aleph K_\aleph(\CA,\CB), ~\label{eq:multi-kernel} 
\end{equation}
where each $K_\aleph$ corresponds to a distinct model.

This is equivalent to an additive model for a property, similar to the construction of an atom-centered decomposition of the total energy in Eq.~\eqref{eq:energy-decomposition}.
In this case, instead, the property $y$ associated with each structure is written as the sum of contributions $y_\aleph(\CA)$ that are associated with the various kernels $K_\aleph$
\begin{equation}
y(\CA) = \sum_\aleph y_\aleph(\CA) = \sum_{\aleph,\CB} x_{\CB} w_\aleph K_\aleph(\CA,\CB) 
\end{equation}
where $x_{\CB}$ are the kernel regression weights for each of the representative structures $\CB$. 
The weights $w_\aleph$ correspond to the estimated contribution that each model will give to the final property, and can be obtained by cross-validation, or by physical intuition. For instance, in the case of multiple radial cutoffs, it is found that much smaller weights should be associated with long-range kernels, consistent with the fact that distant interactions contribute a small (although often physically relevant) contribution to the total energy~\cite{bart+17sa}.
It should also be noted that, provided that the representations corresponding to the kernels are linearly independent,  Eq.~\eqref{eq:multi-kernel} effectively corresponds to a feature space of increased dimensionality, obtained by concatenating the representations that are -- implicitly or explicitly -- associated with each kernel. 

\section*{Conclusions}

We have laid out a mathematical framework, based on the concept of the atomic density, for building representations of atomic environments that preserve the geometric symmetries, and chemically sensible limits. Coupled with kernel regression, this allows the fitting of complex models of physical properties on the atomic scale, both scalars like interatomic potentials (force fields), and tensors such as multipole moments  and quantum mechanical operators. 
We discuss in general terms how kernel regression can be extended to include a sparse selection of reference structures, and to predict and learn from linear functionals of the target property. 
To leverage the many formal similarities between kernel regression and quantum mechanics, we use a Dirac bra-ket notation to formulate the main results concerning the SOAP representations. This notation also helps making apparent the relationship between SOAP representations and other popular density-based approaches to represent atomic structures. 
The framework can be extended and tuned in many different ways to incorporate insight about the relations between properties, structures and representations.
With physical principles such as symmetry and nearsightedness of interactions at its core, we believe this formulation is ideally suited to provide a unified framework to  machine learn atomic-scale properties.

\onecolumngrid

\definecolor{light}{gray}{0.9} 
\centering\colorbox{light}{
\begin{minipage}{0.95\linewidth}
\setlength{\nomitemsep}{0.2\parsep}
\printnomenclature[2cm]
\end{minipage}}
\end{document}